\UseRawInputEncoding
\documentclass[prd,preprint,showkeys,amsmath,amssymb,]{revtex4}
\usepackage{amsmath}  
\usepackage{amsfonts} 
\usepackage{graphicx} 

\usepackage{bm}
\usepackage{color}

\begin{document}

\title{Are LHCb exotics $T_{c\bar{s}0}(2900)^0$, $T_{c\bar{s}0}(2900)^{++}$ and $\overline{X}_0(2900)$ members of an $SU_F(3)$ ${\bf 6}$-plet?}

\author{V.  Dmitra\v sinovi\' c}
\email{dmitrasin@ipb.ac.rs}
\affiliation{Institute of Physics, Belgrade University, Pregrevica 118, Zemun, \\ P.O.Box 57, 11080 Beograd, Serbia}

\date{\today}

\begin{abstract}
Manifestly exotic scalar resonance $X_0(2900)^0$ 
with minimal quark content of $[\bar{c} \bar{s} u d]$ was reported by LHCb \cite{LHCb:2020pxc} in 2020. More recently LHCb reported \cite{LHCb:2022bkt,LHCb:2022bkt1} discovery of manifestly exotic $T_{c\bar{s}0}(2900)^0$ and $T_{c\bar{s}0}(2900)^{++}$ scalar states, degenerate with the $X_0(2900)$. We argue that these are three of six members of the flavor $SU(3)_F$ symmetry ${\bf 6}$-plet. We predict the  partial widths of $D(2900)^{*0}$, the crucial non-strange missing member of the tetraquark ${\bf 6}$-plet, and discuss the optimal decay channels for its detection. 
\end{abstract}
\maketitle


\noindent{\bf Introduction}
Manifestly exotic scalar resonance $X_0(2900)$ in the $D^- K^+$ channel with mass around 2900 MeV and minimal quark content $[\bar{c} \bar{s} u d]$ was reported in 2020 by LHCb \cite{LHCb:2020pxc}. Its decay width is reported as
$\Gamma(X_0(2900) \to D^- K^+) = 57 \pm 12 \pm 4$ MeV. More recently LHCb reported \cite{LHCb:2022bkt,LHCb:2022bkt1} a discovery of manifestly exotic $T_{c\bar{s}0}(2900)^0$ and $T_{c\bar{s}0}(2900)^{++}$ scalar states, with decay widths
$\Gamma(T_{c\bar{s}0}(2900)^0 \to D_s^+ \pi^-) = 119 \pm 26 \pm 13$ MeV; and $\Gamma(T_{c\bar{s}0}(2900)^{++} \to D_s^+ \pi^+) = 137 \pm 32 \pm 17$ MeV, while noticing
its (manifest) degeneracy with the $X_0(2900)$,  \footnote{Indeed, LHCb \cite{LHCb:2022bkt1} noted that ``The obtained mass of the $T_{c\bar{s}0}$ state is consistent with that of another $0^+$ open-charm tetraquark, the $X_0(2900)([cs\bar{u}\bar{d}])$ state discovered in the $D^+ K^-$ final state [19, 20], but their widths and flavor contents are different.''}, albeit without noticing that this degeneracy is one condition for their membership in the flavor $SU_F(3)$ symmetry ${\bf 6}$-plet. 

In this Letter we argue that the $T_{c\bar{s}0}(2900)^0, T_{c\bar{s}0}(2900)^{++}, \overline{X}_0(2900)$ are but three of six degenerate members of an $SU_F(3)$ symmetry ${\bf 6}$-plet, as suggested in \cite{Dmitrasinovic:2004cu,Dmitrasinovic:2005gc}. The remaining members should be the (isovector) $T_{c\bar{s}0}(2900)^{+}$ and an isodoublet $(D^{*0}(2900), D^{*+}(2900))$ of scalar hidden-strangeness cryptoexotics. All members of this ${\bf 6}$-plet were predicted at the same mass, even in the broken $SU(3)_F$ symmetry case, i.e., with $m_s \neq m_{u/d}$. The agreement of the predicted mass degeneracy with the observed masses $T_{c\bar{s}0}(2900)^{++}$ and $\overline{X}_0(2900)$ is impressive, but the different decay widths may yet provide a challenge, which we address below. 

We remind the reader of the 2005 prediction \cite{Dmitrasinovic:2005gc} of an $SU_F(3)$ symmetry ${\bf 6}$-plet of mass-degenerate tetraquarks with a bare mass of 2725 MeV, in a simple nonrelativistic constitent quark model (NRCQM) with 'tHooft strong-hyperfine interaction. The mass 2725 MeV was predicted by fitting model parameters so as to accommodate the then-new, but now defunct \cite{Swanson:2006st,Chen:2016spr} $D^{*}_{sJ}(2632)$ SELEX state \cite{SELEX:2004drx}. Thus, the ${\bf 6}$-plet mass is now free to be  refitted at 2900 MeV. Taking the common mass of ${\bf 6}$-plet as 2900 MeV, we calculate the 
branching ratios using only model-independent features  such as the $SU_F(3)$ symmetry and two-body phase space. The predicted total width is consistent with the measured ones of $T_{c\bar{s}0}(2900)^{0,++}$, but larger than the observed $\overline{X}_0(2900)$ one, roughly by a factor of two. We briefly discuss a number of open theoretical issues that may influence the widths. 

\begin{figure}[tbp]
\centerline{
\includegraphics[width=0.75\columnwidth,,keepaspectratio]{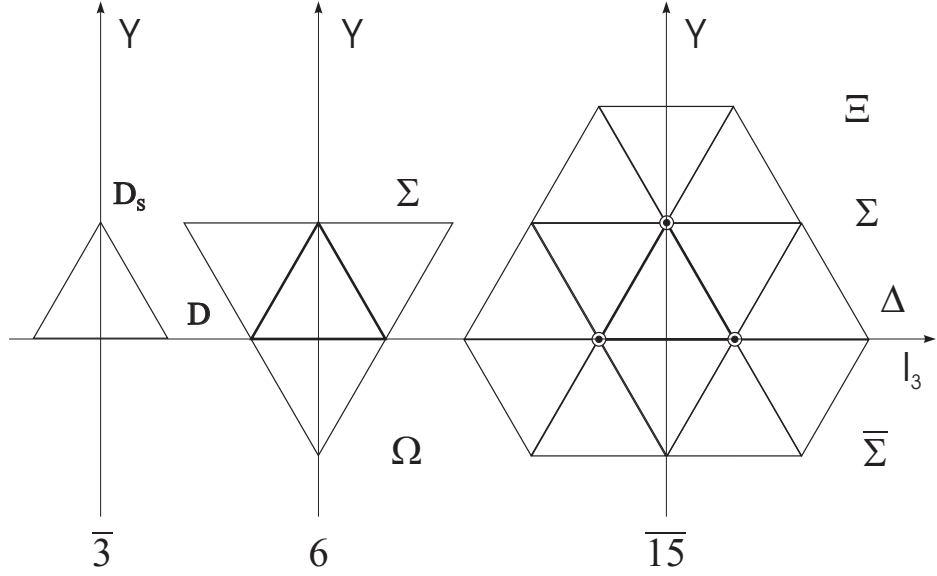}}
\caption{Weight diagrams of SU(3) irreducible representations appearing in the single-charm tetraquark Clebsch-Gordan series.}
\label{f:su(3)}
\end{figure}

\noindent{\bf Masses}
of the ${\bf 6}$-plet 
are subject to certain $SU_F(3)$ flavour symmetry conditions, 
which we derive below from the 
SU(3) flavor wave functions of single-charmed tetraquarks. 
The SU(3) flavor multiplets of C=1 tetraquarks
are given by the flavor SU(3) Clebsch-Gordan series ${\bf 3} \otimes {\bf {\bar
3}} \otimes {\bf {\bar 3}} = {\bf 3} \otimes ({\bf 3} \otimes {\bf
{\bar 6}}) = {\bf {\bar 3}_{\rm A}} \oplus {\bf {\bar 3}_{\rm S}}
\oplus {\bf 6} \oplus {\bf \overline{15}}$. For corresponding SU(3) weight diagrams see Fig. \ref{f:su(3)}.
There are two distinct flavour ${\bf \overline{3}}$-plets in this Clebsch-Gordan series, that are distinguished by their permutational symmetry, or antisymmetry with respect to the interchange of the two quarks:
${\bf 3}_{\rm S, A}$. The two tetraquark anti-triplets (${\bf \bar{3}}$) are analogous to $c{\bar q}$ mesons, which we call cryptoexotics; three members of the sextet (${\bf 6}$) and many of the ${\bf \overline{15}}$-plet do not have $c{\bar q}$ analogons, which makes them exotics. 

The two antisymmetric charmed tetraquark flavour multiplets (${\bf \overline{3}_{A}}$-plet and ${\bf 6}$-plet) have the curious property that all of their members have the same mass\footnote{Note that this degeneracy is a function of the SU(3) symmetry breaking patterns, which, in turn, depend on the the 
strong hyperfine interaction \cite{Dmitrasinovic:2006nx,Dmitrasinovic:2006uk,Yasui:2007dv}.}, in the linear approximation to SU(3) symmetry breaking,  irrespective of their manifest strangeness \cite{Dmitrasinovic:2005gc}, which is straightforward to see  
from their flavor wave functions,
\begin{eqnarray}
|{\rm T_{c\bar{s}q\bar{q}}^{0}} \subset {\bf {6}} \rangle  &=&
\frac{1}{\sqrt{2}}|c
u \left({\bar d}{\bar s} - {\bar s}{\bar d}\right) \rangle
\nonumber \\
|{\rm T_{c\bar{s}q\bar{q}}^{+}} \subset {\bf {6}} \rangle  &=&
\frac{1}{\sqrt{2}}|c d \left({\bar s}{\bar u} - {\bar u}{\bar s} \right) \rangle
\nonumber \\
|{\rm T_{c\bar{s}q\bar{q}}^{++}} \subset {\bf {6}} \rangle  &=&
\frac{1}{2}|c \left(u({\bar u}{\bar s} - {\bar s}{\bar u}) +
d({\bar d}{\bar s} - {\bar s}{\bar d}) \right) \rangle
\nonumber \\
|{\rm D_{}^{0*}} \subset {\bf {6}} \rangle  &=&
\frac{1}{2}|c \left(s({\bar u}{\bar s} - {\bar s}{\bar u}) +
d({\bar d}{\bar u} - {\bar u}{\bar d}) \right) \rangle
\nonumber \\
|{\rm D}^{+*} \subset {\bf {6}}\rangle  &=&
\frac{1}{2}|c \left(s({\bar d}{\bar s} - {\bar s}{\bar d}) +
u({\bar d}{\bar u} - {\bar u}{\bar d}) \right) \rangle
\nonumber \\
|\overline{\rm X}_{0}^{0} \subset {\bf 6} \rangle &=&
\frac{1}{\sqrt{2}}|c s \left({\bar d}{\bar u} - {\bar u}{\bar d} \right) \rangle ~. \
\label{e:6_plet}
\end{eqnarray}
In Eq. (\ref{e:6_plet}) it can be seen that even the two states $({\rm D}^{0*},  {\rm D}^{+*})$ with zero net strangeness contain an $s {\bar s}$ pair one half of the time, which effectively increases their masses by one strange-up/down quark mass difference $m_s - m_{u/d}$. This property turned out in (surprisingly) good agreement with the measured masses of the $D_{s0}^{+}(2317)$ and $D_{0}(2308)$ mesons
\cite{Belle:2003guh,Belle:2003kup,Belle:2003nsh}. 
Three, ${\rm T_{c\bar{s}q\bar{q}}^{0}}(2900), {\rm T_{c\bar{s}q\bar{q}}^{++}}(2900)$ and $\overline{\rm X}_{0}^{+}(2900)$, of the six presumed members of the ${\bf 6}$-plet have been discovered with degenerate masses, within combined uncertainties, so it (only) remains to be seen if this degeneracy will equally hold true for the remaining  strangeness-zero isodoublet $({\rm D}^{0*},  {\rm D}^{+*})$?

This situation closely resembles that of the lowest-mass spin-$\frac{3}{2}$ baryon ${\bf 10}$-plet before the discovery of $\Omega^-(1670)$. There, also, the $SU_F(3)$ symmetry breaking patterns led Gell-Mann \cite{Gell-Mann:1962yej} to the (spectacular) prediction of the mass of the previously missing hyperon, $\Omega^-(1670)$. 

Of course, these tetraquark states have substantial decay widths which may influence their ``dressed'' masses. Next, we examine the observed decay widths, so as to see if they can be used to predict the widths of the as yet undiscovered states?




\noindent{\bf Decay widths} LHCb announced \cite{LHCb:2022bkt,LHCb:2022bkt1} the results of a search for exotic isovector $c\bar{s}$ states, as 
two new resonances with masses of
\[T_{c{\bar s}(q{\bar q})_{I=1}}^{0,++}(2900): M = 2.908 \pm 0.011 \pm 0.020~{\rm GeV}\] and widths of
\[T_{c{\bar s}(q{\bar q})_{I=1}}^{0,++}(2900):\Gamma(T_{c\bar{s}} \to \pi^{\pm} D_s^+) = 136 \pm 23 \pm 11~{\rm MeV}\]
In the $D^- K^+$ channel, on the other hand, there  are two resonances \cite{LHCb:2020pxc}, both described with Breit-Wigner line shapes, the scalar ($J^P = 0^+$) one with parameters
\[X_0(2900): M = 2.866 \pm 0.007 \pm 0.002~{\rm GeV}/c^2 ;\]
\[\Gamma(X_0 \to K^+ D^-) = 57 \pm 12 \pm 4~{\rm MeV};\]
which is roughly two times smaller than the width of $T_{c{\bar s}(q{\bar q})_{I=1}}^{0,++}(2900)$.
Moreover, this is substantially smaller than the expected total width ($\geq 300$ MeV) of such tetraquarks. Can one understand these differences? 

First, note that $T_{c{\bar s}(q{\bar q})_{I=1}}^{++}(2900)$ need not decay only into $D_{s}^+ \pi^+$,
but may also decay into $D_{}^{+} K^+$, which has not been observed (as yet), with only a slightly smaller phase space. 
Similarly, $T_{c{\bar s}(q{\bar q})_{I=1}}^{0}(2900)$ need not decay only into $D_{s}^+ \pi^-$,
but may also decay into $D^{0} K^0$, which has not been observed due to the neutrality of decay products. 
Finally, $X_0 \to K^+ D^-$ is not the only allowed mode of decay - $X_0(2900) \to \overline{D}^{0} \overline{K}^0$ is also allowed. We must therefore examine all these decay widths.

\noindent{\it Two-body decay widths} are given by
\[\Gamma(M \to m_1+m_2) = \frac{1}{8 \pi}
|{\cal M}|^2 \frac{\lambda(M, m_1, m_2)}{M^2}\]
where $|{\cal M}|^2 = f^2_{SU(3)} |{\rm m}|^2$ is the quantum mechanical decay amplitude squared, which factors into $f^2_{SU(3)}$, the squared SU(3) ``isoscalar factor'' in the given specific flavor channel and the  flavor-independent decay amplitude squared $|{\rm m}|^2$, and
\begin{eqnarray}
&& \lambda(M, m_1, m_2) = |{\bf p}_1| = |{\bf p}_2| 
\nonumber \\
&=& \frac{\sqrt{\left(M^2-(m_1-m_2)^2\right) \left(M^2-(m_1+m_2)^2\right)}}{2 M}
\label{e:Mandelstam} \ \end{eqnarray}
is the phase space factor
(also known as the Mandelstam function).
The flavor-independent decay amplitudes ${\rm m}$
are equal, for equal masses $M, m_1,m_2$, to linear approximation.

This circumstance allows us to calculate the ratios of partial widths as
\[\frac{\Gamma(T\to D_{s}^+ \pi^+)}{\Gamma(T\to D_{}^{+} K^-)} =
\frac{f^2_{SU(3)}(T, D_{s}^{+}, \pi^-) \lambda (T, D_{s}^{+}, \pi^-)}{f^2_{SU(3)}(T, D_{}^{+}, K^-) \lambda (T, D_{}^{+}, K^-)} .\] 
provided we are given 
\noindent{\it SU(3) isoscalar factors}
which are just the flavor SU(3) symmetry off-diagonal matrix elements $f_{SU(3)}(T_{c}(2900) \to f) = \langle f | T_{c}(2900) \rangle $. We use the ${\bf 6}$-plet tetraquark flavor wave functions, Eq. (\ref{e:6_plet}), 
and $K^- = s\bar{u}$, $\bar{K}^0 = s\bar{d}$, $\pi^0 = \frac{1}{\sqrt{2}} (u \bar{u} - d \bar{d})$, $\pi^+ = u \bar{d}$, $\eta^{8} = \frac{1}{\sqrt{6}} (u \bar{u} + d \bar{d} - 2 s \bar{s})$, $\eta^{0} = \frac{1}{\sqrt{3}} (u \bar{u} + d \bar{d} + s \bar{s})$, $K^+ = u\bar{s}$, ${K}^0 = d\bar{s}$, $D_{}^{0} = c \bar{u}$, $D_{}^{+} = c \bar{d}$, $D_s^+ = c \bar{s}$. 

\noindent{\it SU(3) isoscalar factors of observed tetraquarks}, which may proceed in (at least) two different channels, e.g.
${\rm T_{c\bar{s}q\bar{q}}^{0}} \to \pi^- D_s^+$, ${\rm T_{c\bar{s}q\bar{q}}^{0}} \to K^0 D^0$.
Their SU(3) isoscalar factors are
\[f(T_{c\bar{s}q\bar{q}}^{0} \to \pi^- D_s^+) = 1/\sqrt{2}\]
\[f(T_{c\bar{s}q\bar{q}}^{0} \to K^0 D^0) = 1/\sqrt{2}\]
The kaonic decay channel is difficult to detect due to two neutrals in the final state. 

Similarly, the $T_{c{\bar s}(q{\bar q})_{I=1}}^{++}(2900)$ SU(3) isoscalar factors are $f(T_{c\bar{s}q\bar{q}}^{++} \to \pi^+ D_s^+) = 1/\sqrt{2}$, $f(T_{c\bar{s}q\bar{q}}^{++} \to K^+ D^+) = 1/\sqrt{2}$
and finally
$f(\bar{X}_0(2900) \to D_{}^{+} K^-) = 1/\sqrt{2}$, $f(\bar{X}_0(2900) \to D_{}^{0} K^0) = 1/\sqrt{2}$.

\noindent{\it SU(3) isoscalar factors of unobserved tetraquarks}
The isodoublet states $(D^{*}(2900)^0, D^{*}(2900)^+)$ decay into more than two flavor channels:
$D^{*}(2900)^0 \to D^0 \pi^0$, $D^{*}(2900)^0 \to D^+ \pi^-$, $D^{*}(2900)^0 \to D^0 \eta$, $D^{*}(2900)^0 \to D_s^+ K^-$, with SU(3) isoscalar factors shown in Table \ref{tab:D_0}. Some of these are eminently observable, e.g., both in the pion ($D^{*}(2900)^0 \to D^+ \pi^-$) and in the kaon ($D^{*}(2900)^0 \to D_s^+ K^-$) channels, due to the charged decay products. These channels should be prime candidates for an experimental search.
\begin{table}
\begin{center} 
\caption{Numerical values of SU(3) matrix elements 
$\langle f  | D^{*0}(2900) \rangle$ for various final states $\langle f |$. Here $\eta^8$ and $\eta^0$ denote the eighth member of the octet and the SU(3) singlet, respectively.}
\begin{tabular}{c@{\hskip 0.1in}c@{\hskip 0.1in}c@{\hskip 0.1in}c@{\hskip 0.1in}c@{\hskip 0.1in}c@{\hskip 0.1in}}
\hline \hline 
\setlength
\hline
\hline
& $\langle D^+ \pi^-|$ & $\langle D^0 \pi^0|$ & $\langle D^0 \eta^{8}|$ & $\langle D^0 \eta^{0}|$ & $\langle D_s^+ K^{-}|$ \\ 
\hline
\hline
$| D^{*0}(2900) \rangle$ & $\frac{1}{2}$ & $\frac{-1}{2\sqrt{2}}$ & $\frac{-1}{2 \sqrt{6}}$ & $\frac{1}{\sqrt{3}}$ & $\frac{-1}{2\sqrt{2}} $ \\
\hline
\hline
\end{tabular}
\label{tab:D_0}
\end{center}
\end{table}

\noindent{\it Decays of observed tetraquarks}
A quick calculation using Eq. (\ref{e:Mandelstam}) and the meson masses from Particle Data Group (PDG) \cite{PDG:2022}, shows that 
the differences in phase space range from 4\% to 28 \%, which cannot account for the large differences, by a factor of two, in the widths of $T_{c\bar{s}}$ and $X_0$. That also shows that the unobserved neutral channel decay ${\rm T_{c\bar{s}q\bar{q}}^{0}} \to K^0 D^0 $ carries approximately one half of the total width. Similarly, the 
$T_{c{\bar s}(q{\bar q})_{I=1}}^{++}(2900)
\to D_{}^{+} K^+$ decay, which is observable in principle, but has not been reported as yet, carries about 50 \% of the total width, which is therefore at least 300 MeV, in agreement with expectations. The $\overline{X}_0(2900) \to D_{}^{0} K^0$ decay, which is difficult to detect on account of the final products neutrality, also carries about 50 \% of the total width. Thus we see that the SU(3) isoscalar factors and phase space considerations do not lead to the convergence of the two decay widths $T_{c\bar{s}q\bar{q}}^{++}$ and $\overline{X}_0(2900)$. Nevertheless, this disagreement should not discourage us too much, as it is comparatively smaller than in the case of the spin-$\frac{3}{2}$ baryon ${\bf 10}$-plet, 
see the Comments below. 

\noindent{\it Predicted widths of unobserved states} The absolute widths of the cryptoexotic $D$ mesons can be calculated from the (known) phase-space factors and the isoscalar factors and one measured 
tetraquark decay width.

1) We predict the ratio of $D^{*}(2900)^0 \to D^+ \pi^-$ and $D^{*0}(2900) \to D_s^+ K^-$ partial decay widths as  
\[\frac{\Gamma(D^{*}(2900)^0 \to D^+ \pi^-)}{\Gamma(D^{*0}(2900) \to D_s^+ K^-)} = {2} \times 1.28 = 2.56 ,\]
Using the $\Gamma(T_{c\bar{s}0}(2900)^{0} \to \pi^- D_s^+)$ decay width to set the total width, we find 
\[\Gamma(D^{*}(2900)^0 \to D^+ \pi^-) 
\simeq 74 \pm 19 ~{\rm MeV}\]
and 
\[\Gamma(D^{*}(2900)^0 \to D_s^+ K^-) 
\simeq 29 \pm 7 ~{\rm MeV}\]
Using the $\Gamma(X_0 \to D^+ K^-)$ decay width to set the total width, we find 
\[\Gamma(D^{*}(2900)^0 \to D^+ \pi^-) 
\simeq 25 \pm 7 ~{\rm MeV}\]
and 
\[\Gamma(D^{*}(2900)^0 \to D_s^+ K^-) 
\simeq 17 \pm 5 ~{\rm MeV}, \]
which puts them within the realm of the measurable. 

2) The decay widths 
of the charged member of the isodoublet $D^{*}(2900)^+$:
$\Gamma(D^{*}(2900)^+ \to D^+ \pi^0)$, $\Gamma(D^{*}(2900)^+ \to D^0 \pi^+)$, $\Gamma(D^{*}(2900)^+ \to D^+ \eta)$, $\Gamma(D^{*}(2900)^+ \to D_s^+ \bar{K}^0)$ are comparable to those of the neutral state, but not easily measurable. Here, again, the trouble is that one of the two decay products is always neutral, so we do not anticipate detection in the foreseeable future.

3) The single-charge isovector tetraquark's $T_{c\bar{s}0}(2900)^{+}$
widths $\Gamma(T_{c\bar{s}0}(2900)^{+} \to \pi^0 D_s^+)$, $\Gamma(T_{c\bar{s}0}(2900)^{+} \to K^+ D^0)$ are comparable to those of $T_{c\bar{s}0}(2900)^{++}$ and $T_{c\bar{s}0}(2900)^{0}$. Again, the main obstruction to an experimental search is that one of its two decay products is always neutral, which makes it unlikely to be observed in the near run.

\noindent{\bf Comments}
The observed common decay width of $\Gamma(T_{c\bar{s}0}^{0,++}(2900)) = 136 \pm 34$ MeV is (at least) two times smaller than naively expected, however. The observed width is only one half of the total width, however, the other half going into (unobserved) decays with at least one neutral object in the final state, which ought to settle the issue of the total width. The same holds for the observed and total widths of $X_0(2900)$, the latter  still being approximately two times too small. That discrepancy should not unsettle us as analogous discrepancies in the lowest-mass spin-$\frac{3}{2}$ baryon ${\bf 10}$-plet are comparatively larger. Nevertheless, the (resonance-peak) Breit-Wigner masses of mass spin-$\frac{3}{2}$ baryons
closely correspond with the bare masses in the flavour SU(3) symmetry schemes \cite{Gell-Mann:1962yej,Oakes:1963zz} and in the constituent quark model \cite{Close:1979}.  
Complicated 
production and decay mechanisms have been invoked for the calculation of each hyperon's individual width \cite{PDG:2022}. Something similar may be expected for scalar tetraquarks
\footnote{The fact that the NRCQM predictions are for bare states is important, as it means that there are no decays (zero width) in that approximation. The physical exotics' (bare and/or dressed) masses lie far above their corresponding two-body thresholds, and therefore should be very wide, with only a lower bound on the width, $\Gamma > 350$ MeV, due to the ``fall-apart'' nature of the decay, much like the $\sigma(500)$ and $K_0^*(700)=\kappa$ light-flavour scalar states, see  Ref. \cite{PDG:2022}.
Opening up of decay channels (``unitarization'' of the calculation) would change not only the decay width, but also the dressed mass of the state. Resonant states could be dynamically generated, see
``Review of scalar mesons'' in Ref. 
\cite{PDG:2022}, i.e., they need not have a bare quark-state ``seed'' (or a Castillejo-Dalitz-Dyson (CDD) pole \cite{Castillejo:1955ed}) at all. Indeed, the $\Delta$ resonance was predicted as a state in a meson-nucleon model, without quarks, and only later confirmed by Chew and Low \cite{Chew:1955zz} in a ``booststrap'' kind of calculation. In the same vein, H.-X. Chen {\it et al.} \cite{Chen:2020aos} have recently argued that ``$X_0(2900)$ can be interpreted as the S-wave $D^{∗-} K^{∗+}$ molecular state''. Even if the as yet unobserved states turn out according to predictions, we shall not know their dynamical origin without further tests.}, as well. 

\noindent{\bf Conclusions }
In this Letter we suggested a simple test of the conjecture that the isotriplet states $T_{c\bar{s}0}^{0,++}(2900)$ and isosinglet $\bar{X}_0(2900)$ belong to an $SU(3)_F$ symmetry ${\bf 6}$-plet. An isodoublet of non-strange cryptoexotic states, ($D_{c({\bar u}q{\bar q})_{I=1/2}}^{0}$, $D_{c({\bar d}q{\bar q})_{I=1/2}}^{+}$) must exist with the (same) mass around 2900 MeV. The neutral member $D^{*0}(2900)$ of this isodoublet ought to also decay into two charged particle channels, $\pi^- D^+$ and $K^- D_s^+$, which should allow ready detection. The observable partial widths $\Gamma(D^{*0}(2900))$ ought to be at most one half of the corresponding partial widths of the  isotriplet $T_{c\bar{s}0}^{0}(2900)$, and/or of the isosinglet $X_0(2900)$. 
This is a rather weak constraint, due to the factor two  difference(s) between the decay widths of  $T_{c\bar{s}0}^{0}(2900)$ and $X_0(2900)$. The prospective discovery of a neutral resonance $D^{*0}(2900)$ at, or near its predicted mass of 2900 MeV, would constitute an unassailable proof of its being a member of a ${\bf 6}$-plet, and therefore of the ${\bf 6}$-plet as a whole, vagaries about decay widths notwithstanding.

Several exotic tetraquarks have been discovered experimentally and discussed theoretically before the latest LHCb batch \cite{LHCb:2020pxc,LHCb:2022bkt,LHCb:2022bkt1} - see e.g. the reviews \cite{Karliner:2017qhf,Liu:2019zoy}, yet there has not been a single instance, to our knowledge, of an exotic discovery following a 
prediction. Terasaki \cite{Terasaki:2003qa} predicted manifestly exotic (double-charged) isovector partners of the 
$D_{s}^{+}(2317)$, which were searched for by Belle \cite{Belle:2015glz} in a narrow strip ($\pm$ 33 MeV) around 2317 MeV, and by BaBar \cite{BaBar:2006} up to 2600 MeV invariant mass, both without success.

Perhaps the only successful prediction of $D_{0}(2308)$ in 2004, Ref. \cite{Dmitrasinovic:2004cu}, led to other specific mass predictions, {\it viz.} that of an isotriplet of $c{\bar s}q{\bar q}$ states, and of  an isoscalar $cs {\bar q}{\bar q}$ state, all belonging to a flavor SU(3)$_F$ symmetry ${\bf 6}$-plet, at 2724 MeV in \cite{Dmitrasinovic:2005gc}. After removing the SELEX constraint, the ${\bf 6}$-plet mass can now be raised to 2900 MeV. 
Refs. \cite{Dmitrasinovic:2004cu,Dmitrasinovic:2005gc,Yasui:2007dv,Dmitrasinovic:2012zz} also predicted existence of a new ${\bf \overline{15}}$-plet of tetraquarks, but little  can be said about their masses with any certainty. 

All of the exotics observed thus far have good isospin, yet their (prospective) SU(3) labels are rarely, if ever discussed. Here we suggested, perhaps for the first time, a specific test of an exotic charmed tetraquarks' new SU(3) multiplet. It appears desirable to put this suggestion to an experimental test.


\noindent{\bf Acknowledgments}
The author acknowledges 
informative correspondence
with Drs. Yinrui Liu and 
Ma Ruiting of the LHCb Collaboration. This research was funded by the Serbian Ministry of Education Science and Technological Development, grant number 451-03-68/2020-14/200024.


\end{document}